\documentclass[aps,prx,twocolumn,showpacs,superscriptaddress,longbibliography]{revtex4-1}

\usepackage{amssymb}
\usepackage{amsfonts}
\usepackage{amsmath}
\usepackage{bm}
\usepackage{textcomp}
\usepackage{color}
\usepackage{longtable}
\usepackage{graphicx}
\usepackage{dcolumn}
\usepackage[a4paper=true,pagebackref=false]{hyperref}
\usepackage[normalem]{ulem}
\usepackage{physics}

\usepackage{xr}

\makeatletter

\newcommand*{\addFileDependency}[1]{
\typeout{(#1)}
\@addtofilelist{#1}
\IfFileExists{#1}{}{\typeout{No file #1.}}
}\makeatother

\newcommand*{\myexternaldocument}[1]{%
\externaldocument{#1}%
\addFileDependency{#1.tex}%
\addFileDependency{#1.aux}%
}

\myexternaldocument{./supplement/main}

\begin{document}


\title{Probing confined vortices with a superconducting nanobridge}

\author{M.~Foltyn}
\email{foltyn@ifpan.edu.pl}
\affiliation{Institute of Physics, Polish Academy of Sciences, Aleja Lotnikow 32/46, PL 02668 Warsaw, Poland}

\author{K.~Norowski}
\affiliation{Institute of Physics, Polish Academy of Sciences, Aleja Lotnikow 32/46, PL 02668 Warsaw, Poland}

\author{M.J.~Wyszy\'nski}
\affiliation{Department of Physics $\&$ NANOlab Center of Excellence, University of Antwerp, Groenenborgerlaan 171, B-2020 Antwerp, Belgium}

\author{A.S.~de~Arruda}
\affiliation{Instituto de F\'isica, Universidade Federal de Mato Grosso, 78060-900 Cuiab\'a, Mato Grosso, Brazil}

\author{M.V.~Milo\v{s}evi\'{c}}
\email{milorad.milosevic@uantwerpen.be}
\affiliation{Department of Physics $\&$ NANOlab Center of Excellence, University of Antwerp, Groenenborgerlaan 171, B-2020 Antwerp, Belgium}
\affiliation{Instituto de F\'isica, Universidade Federal de Mato Grosso, 78060-900 Cuiab\'a, Mato Grosso, Brazil}

\author{M.~Zgirski}
\email{zgirski@ifpan.edu.pl}
\affiliation{Institute of Physics, Polish Academy of Sciences, Aleja Lotnikow 32/46, PL 02668 Warsaw, Poland}

\date{\today}

\begin{abstract}
We realize a superconducting nanodevice in which vortex traps in the form of an aluminum square are integrated with a Dayem nanobridge. We perform field-cooling of the traps arriving to different vortex configurations, dependent on the applied magnetic field, to demonstrate that the switching current of the bridge is highly sensitive to the presence and location of vortices in the trap. Our measurements exhibit unprecedented precision and ability to detect the first and successive vortex entries into all fabricated traps, from few hundred nm to 2 $\mu$m in size. The experimental results are corroborated by Ginzburg-Landau simulations, which reveal the subtle yet crucial changes in the density of the superconducting condensate in the vicinity of the bridge with every additional vortex entry and relocation inside the trap. An ease of integration and simplicity make our design a convenient platform for studying dynamics of vortices in strongly confining geometries, involving a promise to manipulate vortex states electronically with simultaneous \textit{in situ} control and monitoring.
\end{abstract}
\maketitle

\begin{figure}[t]
\centering
\includegraphics[width=0.48\textwidth]{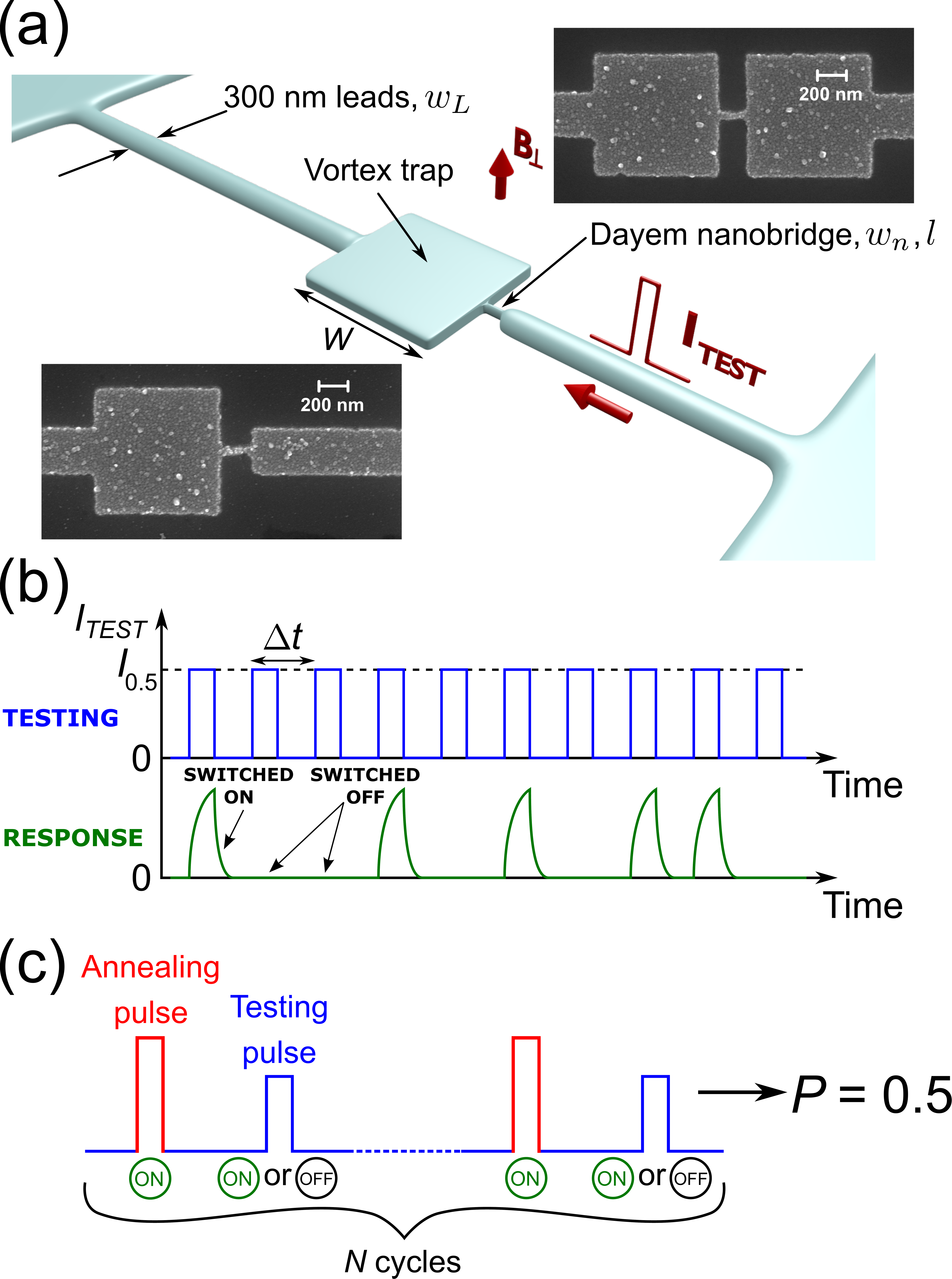}
\caption{\label{fig:Intro}The sample layout and the measurement protocol. (a) The sample consists of the square vortex traps connected via a short Dayem nanobridge. The traps are connected to the contact pads through $\sim100~\mu$m-long and 300-nm wide leads. Electric signal is applied to one of the contact pads, while the other pad is grounded. Insets show SEM images of the typical aluminum structures of the single and double vortex traps. (b) We send a train of $N$ pulses to the nanobridge in order to determine its switching probability $P$. For a certain current amplitude of the testing pulses ($I_{0.5}$) the nanobridge switches to the normal state around $N$/2 times. (c) In each cycle we send additional pulse before every testing of the nanobridge, which acts as a “reset”. The annealing pulse always switches the nanobridge to the normal state and overheats the vortex trap above $T_{c}$. In our experiment the real probing pulse is more complex and consists of a short part of higher amplitude intended to test the junction (so-called \textit{testing pulse}), and much longer, typically 5$~\mu$s long, sustain part, enabling the read-out of the state of the bridge with a low-pass-filtered line.}
\end{figure}

\section{Introduction}
The London prediction of the flux quantization in superconductors was verified in seminal experiments of Deaver and Fairbank~\cite{Deaver1961}, and Little and Parks~\cite{Little1962} performed on superconducting cylinders. Subsequently it was shown that such microscopic tubes of magnetic flux in superconductors can arrange themselves into a regular lattice, proving the earlier theoretical finding of Abrikosov~\cite{Abrikosov1957}. Creation of vortices in a superconducting material enables coexistence of normal inclusions in otherwise superconducting environment, and allows the system to retain superconducting properties up to larger applied magnetic fields.

The behavior of vortices in mesoscopically engineered superconductors, with key geometric features comparable in size to the governing length scales of superconductivity (coherence length $\xi$ and/or penetration depth $\lambda$), has been of tremendous research interest since the 1990's \cite{Moshchalkov1995,Geim1997,Schweigert1998,Peeters1999,Geim2000}. Only since recently, the devices based on single-vortex control have been proposed and realized \cite{Krasnov2015}, partially owing to the prospect of surprisingly high speed of vortex `writing' and `erasing' \cite{Embon2017,Vodolazov2020}. Direct imaging of vortices was realized to date by numerous microscopy techniques, involving scanning tunnel microscopy~\cite{Waszczak1989,Timmermans2016}, Lorentz microscopy~\cite{Harada1996}, differential magneto-optical technique~\cite{Soibel2000}, magnetic decoration imaging~\cite{Fasano1999}, scanning Hall probe microscopy~\cite{Martinis2004}, or scanning superconducting quantum interference device (SQUID) microscopy~\cite{Koshnick2008, Zeldov2013, Hilgenkamp2015}. The ballistic Hall magnetometry on mesoscopic superconductors could capture magnetization jumps caused by the single vortex entry~\cite{Geim1997}. Detection of vortex entry was also done indirectly by the measurement of magnetoresistance~\cite{He2011, Peeters2012, Mills2016, Kunchur2016}. Similarly, normal metal-insulator-superconductor junctions were employed to probe different configurations of vortices within small superconducting islands, owing to sensitivity of the quasiparticle tunneling to the Meissner currents in the junction~\cite{Kadowaki2005,Timmermans2016}. 

The techniques visualizing and sensing distribution of vortices in mesoscopic samples are thus quite broad. However, the approaches based on electrical probing of the vortex configuration which would be compatible with construction of novel-concept devices remain very limited. The ability to manipulate and detect single vortices is an essential requirement for construction of functional superconducting devices in the emerging field of the vortex electronics~\cite{Krasnov2015}. For example, it was demonstrated that a controlled placement of a vortex can strongly influence the characteristics of an adjacent Josephson junction~\cite{Berdiyorov2011,Krasnov2019}, the use of which in devices is at the very core of the modern quantum technology.

In this paper, we use a Dayem nanobridge connected to aluminum nanosquare traps as a sensing element for the vortex configurations in the traps. We show that the entry and arrangement of vortices in the field-cooled sample directly influence the pattern of the switching current of the nanobridge. The demonstrated implementation of the device allows electronic detection of individual vortices, which can also be treated as a means to control the transport properties of the bridge. These features make our device a convenient platform for presenting desired functionalities rooted in the physics of the superconducting vortices. 

\vspace{1.5cm}

\section{Methodology}
\subsection{Samples}
We fabricate a series of samples by means of single-step standard electron-beam lithography. To control the position of vortices reliably it is often required to embed the pinning centers in the studied devices~\cite{Krasnov2019}. It is also possible to confine their position by providing a suitable geometry~\cite{Geim1997,Baelus2000}, which is the strategy employed in our work. The structures are 30-nm thick aluminum layers evaporated on the silicon substrate, consisting of vortex traps in a form of squares linked by a narrow Dayem nanobridge [typically approximately equal to$50\times200$ nm, cf. Fig.~\ref{fig:Intro}(a)]. The connecting leads have much smaller width than the vortex box (typically approximately equal $300$~nm, cf. Table~\ref{fig:Parameters}). Such a design ensures that in the studied range of magnetic fields vortex will penetrate the nanostructure only within the trap area(s). The sides of the fabricated vortex boxes were between 580~nm and 1.9$~\mu$m. Five samples are symmetric, with the same box on each side of the bridge, two samples contain one box only and two are reference nanostripes (contain no boxes).

\begin{table}[h!]
\begin{tabular}{|p{2.0cm}|p{1.0cm}|p{0.8cm}|p{0.8cm}|p{0.8cm}|p{0.8cm}|p{0.8cm}|}
\hline
Sample& $W$ (nm)&$l$ (nm) &$w_n$ (nm) &$w_L$ (nm) & $I_{SW}$ ($\mu$A) & Chip\\
\hline
\textbf{A} (Double)& 860 & 195 & 55 & 310 & 50.7 & 1 \\
\textbf{B} (Single)& 860 & 220 & 45 & 320 & 21.9 & 1 \\
\textbf{C} (Single)& 980 & 200 & 40 & 245 & 22.5 & 2 \\
\textbf{D} (Double)& 1050 & 140 & 40 & 285 & 29.7 & 3 \\
\textbf{E} (Nanowire)& - & 300 & 60 & 540 & 58 & 4 \\
\textbf{F} (Nanowire)& - & 210 & 70 & 600 & 45 & 5 \\
\hline
\end{tabular}
\caption{\label{fig:Parameters}Measured parameters of the selected studied samples. Nanostructures are fabricated in single and double box configurations, with side $W$ [cf. Fig.~\ref{fig:Intro}(a)], or as a nanostripe interrupted with a Dayem bridge of the length $l$ and the width $w_n$ (as extracted from the SEM images). Aluminum thickness for all samples is 30$~$nm. $w_L$ stands for width of the lead nanostripe. Nanobridge switching currents $I_{SW}$ are measured for probability $P=0.5$ at bath temperature $T_0=400\,$mK. Samples A and B are fabricated on the same silicon chip, using the same lithography and evaporation processes. Detailed SEM photos of each sample are provided in Fig. S1 within the Supplemental Material~\cite{Supplemental}.}
\end{table}

\subsection{Measuring protocol}
We measure the switching current of the nanobridges in the dilution refrigerator at constant bath temperature of $T_0=400~$mK. We use the same measurement protocol like in our previous works~\cite{Zgirski2015,Zgirski2020}. The method is based on the testing of nanobridge with a train of $N$ identical current pulses. The pulse may transit the nanobridge to the normal state with temperature above $T_c$ (such transition is called a switching event) or it may remain in the superconducting state. Switching events are recorded on the oscilloscope as voltage pulses [Fig.~\ref{fig:Intro}(b)]. Depending on the pulse amplitude $I_{test}$, the number of switchings in the train $n$ may vary from 0 to $N$, although $n$ takes values different than 0 or $N$ only for very narrow range of pulse amplitudes, rendering S-shaped dependence of the switching probability $P=n/N$ on the testing current. A few iterations in testing current amplitude allow to find switching current $I_{SW}$ corresponding to $P$=0.5 to be found with sufficient accuracy.

We introduce the additional prepulse (i.e. annealing pulse) before each probing pulse, whose amplitude is always higher than $I_{SW}$ [Fig.~\ref{fig:Intro}(c)]. The annealing pulse always switches the nanobridge to the normal state, which leads to overheating of the vortex trap above $T_c=1.3~K$\cite{Zgirski2021}. The subsequent cooling, taking place in the presence of constant magnetic field, is governed by electron-phonon coupling and lowers the temperature of the box below $T_c$ in a few tens of nanoseconds~\cite{Zgirski2018} leading to capturing of the quantized filaments of the magnetic field in the box. The number and configuration of trapped vortices depend on the external magnetic field. Owing to electron-phonon coupling and QP diffusion the sample continues to cool down back to the bath temperature $T_0$ in a few microseconds~\cite{Zgirski2019}. The interval between pulses is much longer than the thermal relaxation time of the sample, typically $\Delta t=100~\mu$s. The Dayem nanobridge with adjacent boxes are well thermalized at the bath temperature $T_0$ when the testing pulse arrives. Hence, our protocol provides thermal cycling of the sample for each prepulse and for some testing pulses, namely those, which are assisted by switching of the nanobridge to the normal state. Of note, the global temperature of the cryostat remains unaltered during the whole experiment, for the Joule heating it involves only the mesoscopic sample, i.e. bridge, vortex box and adjacent nanoleads. The annealing pulse can be perceived as a reset pulse removing any magnetic memory from the system and initializing the state of the boxes. If the testing pulse changes the vortex configuration, the annealing pulse would bring the system to the statistically same initial condition.

\subsection{Validation by Ginzburg-Landau simulations}
To gain insights into vortex states inside the trap(s) and their relation to the switching current of the nanobridge connected to the trap(s), we perform time-dependent Ginzburg-Landau (TDGL) simulations. The behavior of the superconducting condensate in the TDGL approach is described by a complex-valued order parameter which is allowed to vary in time and space, under the effect of applied magnetic field and/or current. We use the TDGL formalism derived for dirty superconductors \cite{kramer, gl_app}, where the dimensionless equation for the order parameter reads:
\begin{equation}
\begin{split}
\frac{u}{\sqrt{1+\gamma^2\abs{\Psi}^2}}\left(\frac{\partial}{\partial t} + i\varphi + \frac{\gamma^2}{2}
\frac{\partial \abs{\Psi}^2}{\partial t}\right)\Psi \\ = 
(\grad - i \vb{A})^2\Psi + \left(1-\abs{\Psi}^2\right)\Psi.
\end{split}
\label{eq:gl1}
\end{equation}
Here $u\approx 5.79$ is the ratio of the relaxation time for the amplitude and phase of the order parameter \cite{kopnin_book}. $\vb{A}$ is the magnetic vector potential due to applied magnetic field, and $\varphi$ is the electrostatic potential. Parameter $\gamma$($=5$) embodies the effect of disorder in the sample, and characterizes the influence of the finite inelastic scattering time on the behavior of the condensate. Eq.~\eqref{eq:gl1} is solved self-consistently with the equation for the scalar electrostatic potential
\begin{equation}
\laplacian{\varphi} = \div \Im\left[\Psi^* (\grad - i \vb{A})\Psi\right].
\label{eq:gl2}
\end{equation}
An external current is applied as a boundary condition to the latter Poisson equation \eqref{eq:gl2}. In these equations, all lengths are measured in units of coherence length $\xi$, magnetic fields in units of the upper critical field $H_{c2}$, time in units of the Ginzburg-Landau time $\tau_{GL} = \pi\hbar / 8k_BT_c u$, current densities in units of $j_0=\frac{3\sqrt{3}}{2}j_{dp}$, where $j_{dp}$ is the depairing current density, and voltage in $\varphi_0 = \hbar / e^* \tau_{GL}$.

The equations are then discretized using the link-variable method, on a Cartesian grid, where the order parameter $\Psi$ is defined on lattice nodes and the vector potential $\vb{A}$ on the links between them. The time step is chosen sufficiently small to guarantee the numerical stability of the forward-time central-space integration scheme for Eq.~\eqref{eq:gl1}. Discretized equations are solved numerically on a lattice with a uniform grid constant of 22.5~nm, which must be taken significantly smaller than the shortest length scale in the system - the coherence length $\xi$ at the considered temperature. The value of $\xi(T=0)$ is obtained through the dirty-limit formula $\xi(0)=0.855\sqrt{\xi_0 l}$, where $\xi_0\approx 1600$ nm is the BCS coherence length, and $l\approx 11.5$ nm is the mean free-path obtained from the experimentally measured normal-state resistivity of the reference samples. Using the well-established temperature dependence of coherence length~\cite{Muller2012} from the two-fluid model $\xi(\tilde{t})=\xi(0)\frac{\sqrt{1-\tilde{t}^4}}{1-\tilde{t}^2}$, where $\tilde{t}=T/T_c$, we obtain $\xi($400~mK$)=128$~nm, which is used in all simulations of the experimental data.

Note that in the performed simulations, no other fitting parameters are used. The geometric parameters are taken as listed in Table~\ref{fig:Parameters}, except for the width of the nanobridge $w_n$ (which is nontrivial to properly extract it from the experimental images due to the granularity of the junction, see Supplemental Material~\cite{Supplemental}). In simulations, $w_n$ is adjusted to an effective uniform width, which yields the same critical current density of nanobridges in all samples in the absence of the magnetic field, following the experimentally measured switching current $I_{SW}$.
\begin{figure*}
\centering
\includegraphics[width=\textwidth]{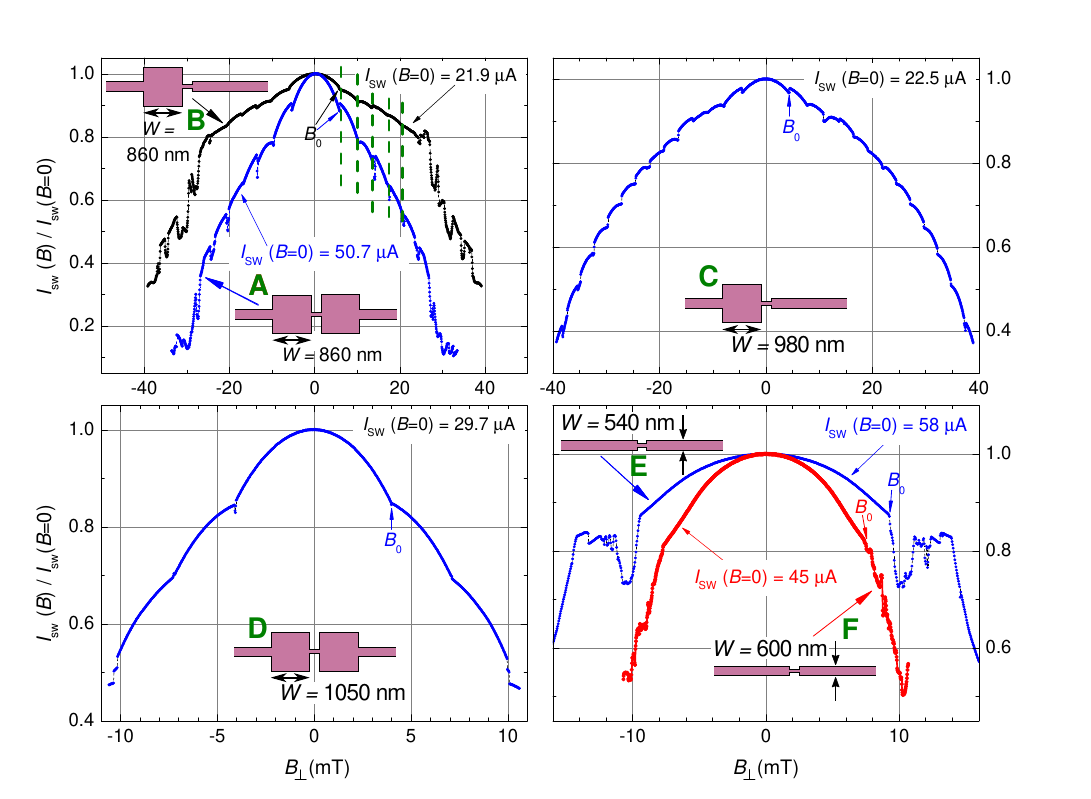}
\caption{\label{fig:ISW}Normalized nanobridge switching currents versus the perpendicular magnetic field $B_{\bot}$ for a series of vortex traps (samples A-D, sizes indicated in the panels) and reference nanostripes (samples E and F), measured at bath temperature $T_0=400$~mK. Up to the first vortex entry $I_{SW}$ decreases monotonously with magnetic field (Meissner state) and changes abruptly at $B_0$. Samples A and B, in double and single trap configuration respectively, both with the side of the trap $W=860$~nm, exhibit the same threshold magnetic field values for successive vortex entries into the trap (marked with vertical dashed lines in the top-left panel) although $I_{SW}$ decreases faster with the field for the sample with a double vortex trap. Nanostripes E and F lack the two-dimensional confining effect on vortices, and above $B_0$ exhibit irregular, but fully reproducible pattern of $I_{SW}(B_{\bot})$.}
\end{figure*}

In order to accurately simulate the experimental protocol, for each value of applied magnetic field an independent current sweep is executed. First, the order parameter is initialized randomly to simulate freezing in from the normal state, resulting in a stochastic initial state. The stochasticity of the nucleation process makes it possible to obtain different (metastable) initial vorticity states at the same magnetic field. Subsequently, the system is evolved deterministically as the applied current is ramped up until the phase-slip event occurred indicating the onset of the switching process.

\begin{figure}
\centering
\includegraphics[width=0.95\linewidth]{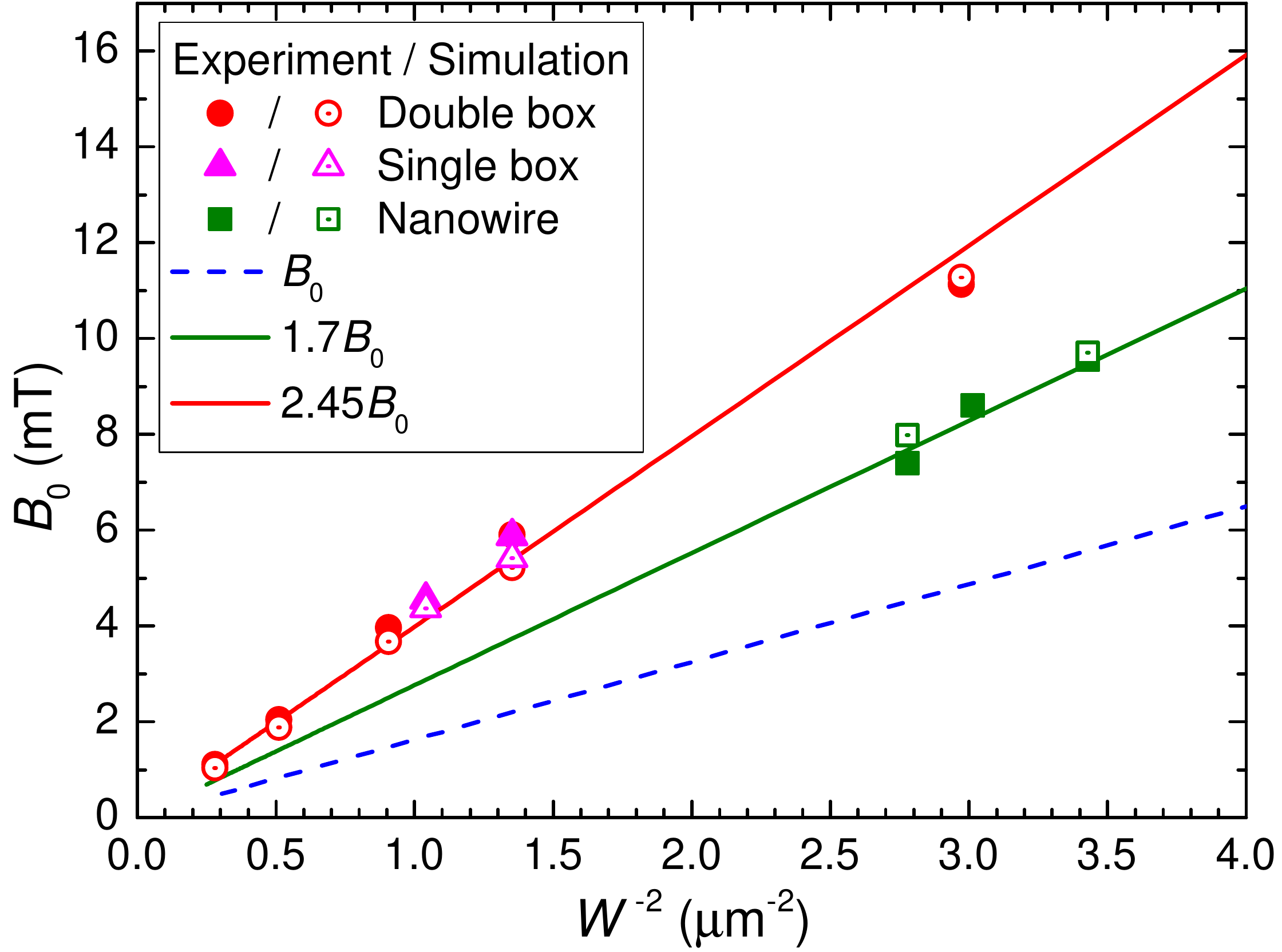}
\caption{\label{fig:Summary}Vortex penetration field $B_0$ for different considered vortex boxes and stripes. Experimental data (full symbols) for single, double boxes and nanowires, extracted from Fig.~\ref{fig:ISW}, as well as the simulated data points (hollow symbols), follow the analytical expression for metastable equilibrium of the vortex in the superconducting strip [Eq.~\eqref{eq:1}], up to a multiplying constant $\beta=2.45$ for traps, and $\beta=1.7$ for stripes. For visual comparison, the dashed line shows the values of $B_0$ for $\beta=1$.}
\end{figure}

\begin{figure*}
\centering
\includegraphics[width=0.85\textwidth]{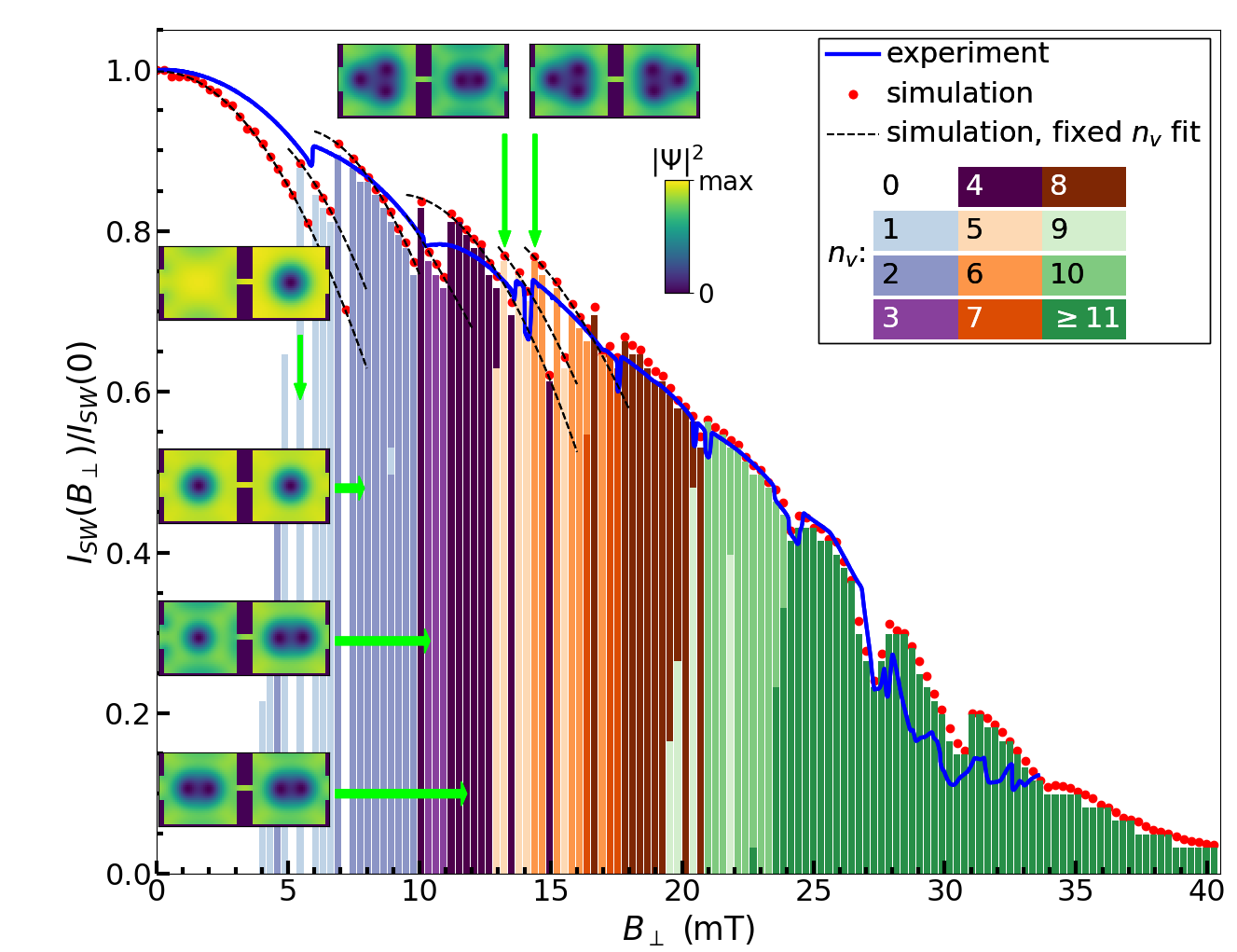}
\caption{\label{fig:DevA}Normalized critical current of sample A, calculated by TDGL simulations (red points). Solid line shows experimental data. Each simulated point is calculated independently for the fixed value of the applied magnetic field. It is obtained as a result of successive equilibrations at step-increasing values of the applied current. The switching current is defined as the one corresponding to the onset of the phase-slip process in the bridge or, at higher fields, in the leads. The color bars denote the total number of trapped vortices $n_v(B_{\bot})$ at the given applied magnetic field. This number is stochastic: it is possible to obtain a different number of trapped vortices when repeating the simulation starting from randomized initial conditions (simulating nucleation from the quasinormal state). The trapping stochasticity is at the root of not sharp transitions between vortex configurations i.e. reentrant behavior is observed whenever we see a decrease in $v(B_{\bot})$ although $B_{\bot}$ is increased. The stochasticity may explain occasional rounding of the experimental curves since they result from measurements performed over $N=1000$ realizations for each field. Noteworthy, the simulated points are grouped according to own vortex number. Those families are marked with broken lines in the plot to provide a convenient guide for an eye. For some magnetic field values, simulation predicts expulsion/entry of the vortex during testing pulse before the switching current of the nanobridge is reached (visible as a two-color bar, where the change of color marks the current for which vortex was expelled or has entered). At a field of approximately $27\,$mT vortices start to enter into leads giving rise to irregular pattern of $I_{SW}(B_\bot)$ (cf. Fig.~\ref{fig:DevE}).}
\end{figure*}

\begin{figure}[b]
\centering
\includegraphics[width=\linewidth]{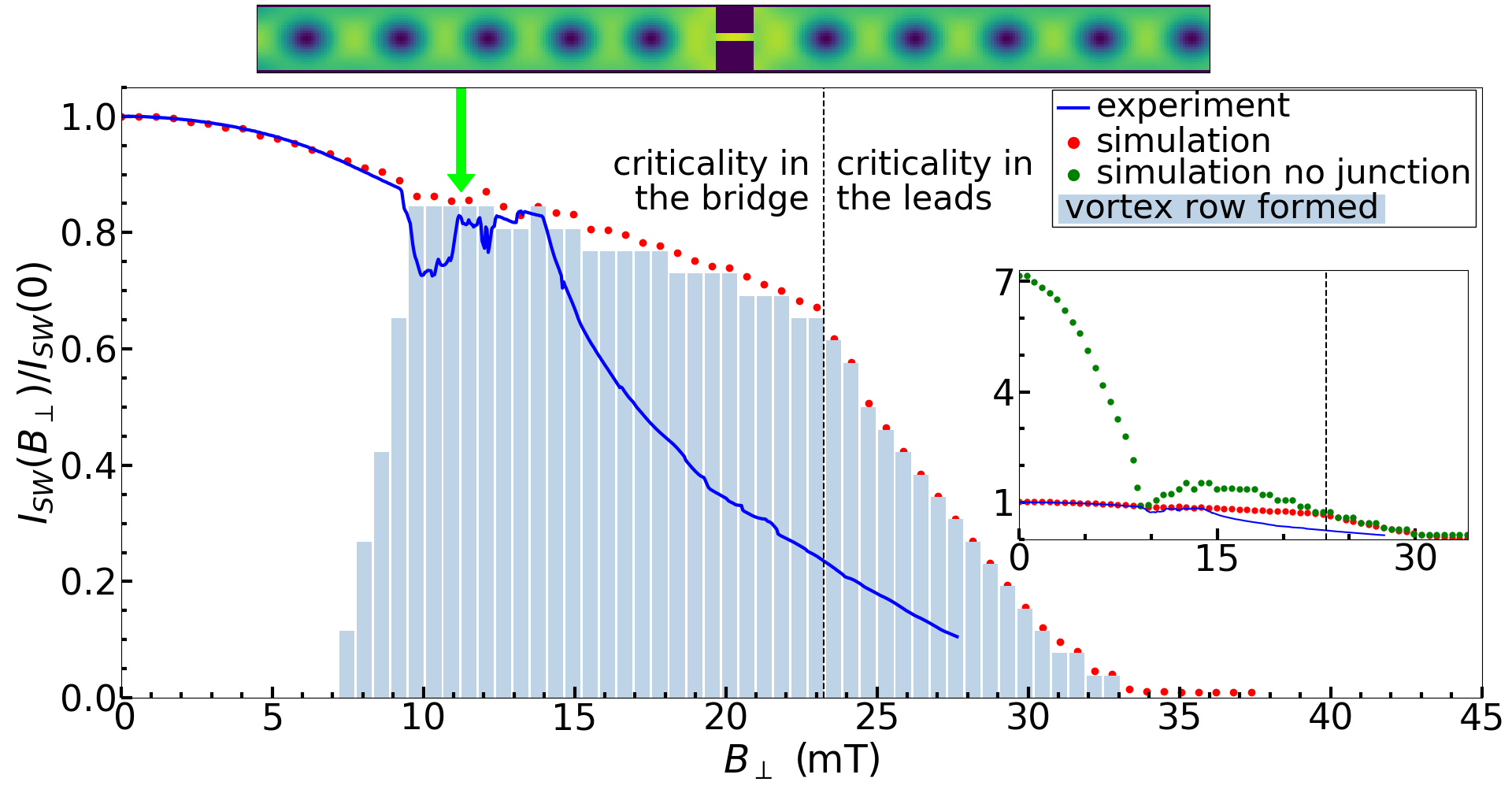}
\caption{\label{fig:DevE}$I_{SW}(B_\bot)$ characteristics of sample E, calculated using TDGL simulations (red dots) in comparison with experimental data (blue line). The bars show the range of fields and currents for which the vortex row exists in the nanowire. At the onset of the vortex entry, the row is unstable at higher applied currents and it is expelled from the sample (such an event is marked as the termination of the bar before the switching current is reached). The inset shows the comparison of the simulation presented in the main panel to the simulated critical current of the nanostripe of the same size, but without any nanobridge. Noteworthy, in the latter case, the $I_{SW}$ shows nonmonotonous behavior attributed to the changes in the edge currents and the order parameter after vortex row enters into the stripe.}
\end{figure}

\begin{figure*}
\centering
\includegraphics[width=0.9\textwidth]{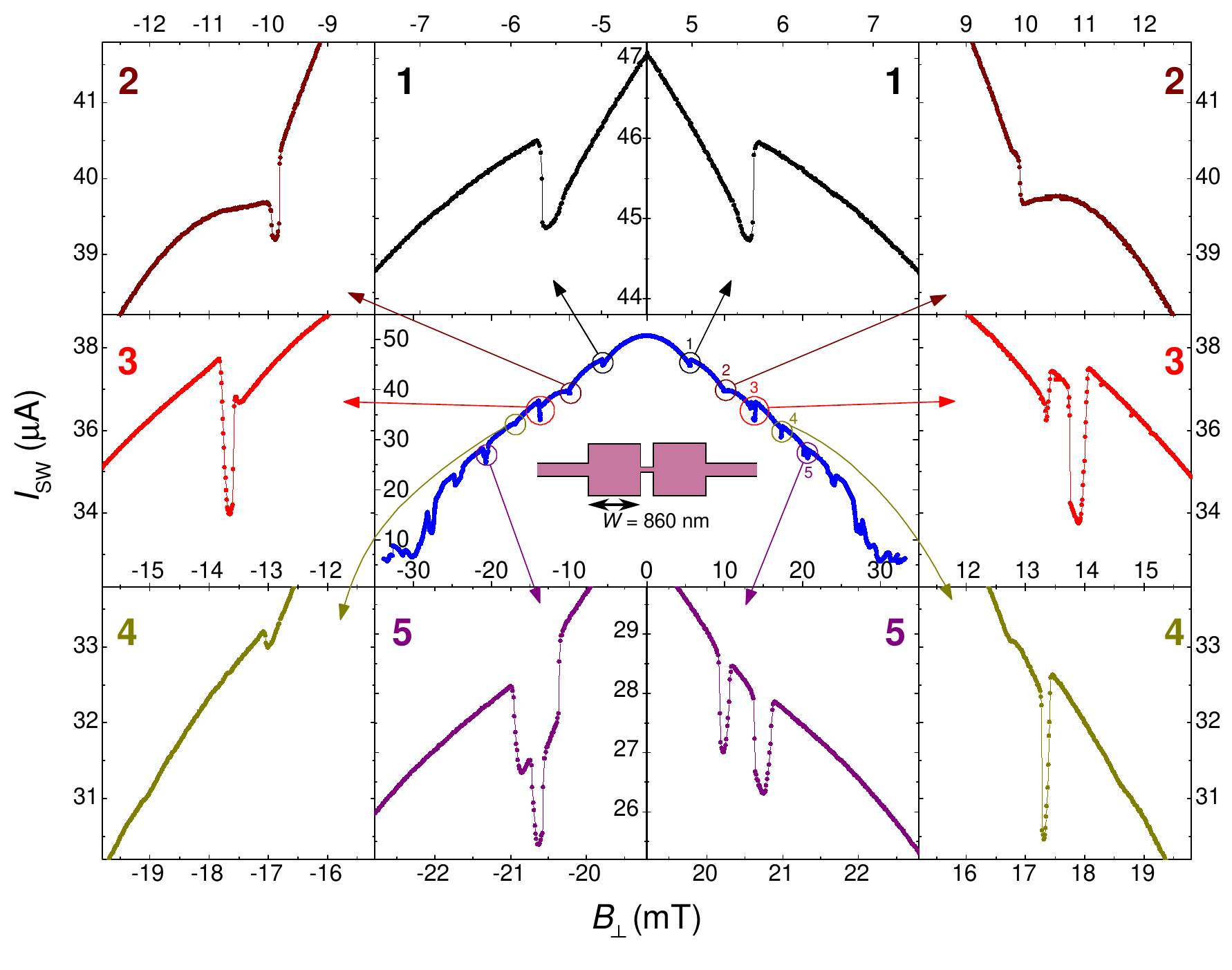}
\caption{\label{fig:Zoom}Switching current versus the perpendicular magnetic field $B_{\bot}$ (sample A) for the nanobridge connecting two vortex traps, both with side of $W$ = 860~nm (see inset), measured at bath temperature $T_0=400~$mK, with enlarged transitions to the next vorticity level.}
\end{figure*}

\section{A Dayem nanobridge as a vortex sensor}
In Fig.~\ref{fig:ISW} we show the collection of $I_{SW}$ measured as a function of magnetic field for various fabricated samples. Independently of the sample geometry, for low applied perpendicular magnetic fields we see monotonous lowering of the $I_{SW}$, associated with depletive effect of Meissner screening currents on the superconducting order parameter along sample edges \cite{Schweigert1998,Peeters1999}. As we increase the magnetic field we observe a nearly periodic series of cusplike changes of the $I_{SW}$ of the nanobridge. We associate the appearance of the steps with the successive entries of vortices, at magnetic fields corresponding to quantized magnetic flux through the vortex traps.

\subsection{Vortex penetration field}

The first observed cusp in $I_{SW}(B_{\bot})$ marks the first vortex penetration field $B_0$. For long type-II superconducting stripes it is expected that upon exceeding $B_0$ the Gibbs free energy exhibits a (meta)stable minimum for existence of vortices in the middle of the strip, separated by Bean-Livingston barriers from the edges of the strip\cite{Bean1964}. This threshold magnetic field for vortex penetration takes the functional form:
\begin{equation}
\label{eq:1}
B_0=\beta B_m=\beta\frac{\pi\Phi_0}{4W^2},
\end{equation}
where $\Phi_0$ is the magnetic quantum flux ($h/2e$), $W$ the width of the strip. $\beta$ is the phenomenological scaling parameter, which in the original prediction for stripes is equal to 1 \cite{Clem1998,Maksimova1998}, if we assume that the local energy minimum in the middle of the strip (but not necessarily the ground state) is sufficient to stabilize the vortex ($B_m$ is the field value for which such a minimum appears for the first time). A more restrictive threshold field $B_G$ for the existence of vortices in the strip was proposed in Ref.~\onlinecite{Likharev1972}. $B_G$ corresponds to the field for which the local energy minimum becomes the ground-state of the system. At fields between $B_m$ and $B_G$, the Gibbs free energy has a local minimum at the center of the strip. The vortex may then leave the strip but only if it overcomes the energy barrier due to thermal activation. In practice, we expect that the vortices become stable in the strip at a field between $B_m$ and $B_G$. These predictions were experimentally validated in Ref.~\onlinecite{Martinis2004}, by direct observation of vortices using Hall-probe microscopy. A corresponding condition was provided in Ref.~\onlinecite{Hilgenkamp2008}. The authors considered the interplay between two mechanisms: the thermal nucleation of vortex-antivortex pairs during field cooling of the strip, and the vortex escape through the energy barrier that exists above $B_m$. Comparison of the two competing effects leads to the critical field for vortex trapping $B_0$, with $\beta=2.1$.

In our experiment we expect to see a similar enhancement of $B_0$ in comparison with $B_m$. For applied fields below $B_0$, the Meissner screening currents are increasing with magnetic field. At each entry of a new vortex these screening currents reduce and the net spatial distribution of the field is changed. Therefore the order parameter of the bridge is affected by both the stray field of the vortex state~\cite{Krasnov2019} and the detailed pattern of the screening currents in the vicinity of the bridge. The first effect has more of a long-range character while the second is more local, within a coherence length from the bridge. Both effects contribute to the fact that nanobridge detects the transition: the $I_{SW}(B_{\bot})$ sharply changes at $B_0$.

We collect the data for all single and double trap geometries as well as for the reference nanostripes, to plot the first-vortex-entry field $B_0$ against $1/W^2$ in Fig.~\ref{fig:Summary}. The functional dependence of Eq.~\eqref{eq:1} describes our data very well, albeit with $\beta=2.45$. Our data for nanostripes yield $\beta=1.7$. The further increase of $\beta$ in boxes, compared to stripes, is expected to come from their small sizes, which are comparable to a penetration depth of the magnetic field $\lambda_L$. Consequently, the sample can remain in the Meissner state up to the higher fields. On the other hand, the vortex field lines close up to a big extent through the borders of the sample, which seems to increase the energetic cost of hosting the vortex inside the trap.

To additionally validate these findings, we employe the TDGL simulations on selected samples. Always starting from the quasinormal state and with 100 repetitions for each value of the applied magnetic field, we increase the field incrementally, in absence of applied current, and recorded the threshold field $B_0$ for which the probability of vortex entry was 50\%. So obtained vortex penetration field is shown together with the experimental data in Fig.~\ref{fig:Summary}.  

\subsection{Behavior of critical current with vortex penetration and ordering}
We next proceed to examine the effect of the device geometry on $I_{SW}(B_{\bot})$. We fabricate two similar samples (A and B), using the exact same process and place them on the same silicon chip. The only difference is that one is mirror symmetric, with one box on each side of the nanobridge [Fig.~\ref{fig:ISW}, sample A, blue curve], while the second sample contains just one box [cf. Fig.~\ref{fig:ISW}, sample B, black curve]. Both $I_{SW}(B_{\bot})$ characteristics show same features - the penetration field of first vortex is nearly identical (yielding flux approximately equal to $2.45\Phi_0$ through each trap, cf. Fig.~\ref{fig:ISW}), and magnetic field values at which successive vortices enter into the trap(s) are the same for both samples [$\Delta B_\bot\approx 3$~mT, i.e. yielding just over one quantum of flux per trap (approximately equal to $1.1\Phi_0$)], up to the applied field of approximately equal to $25$~mT, where $I_{SW}$ shows accelerated decrease and its periodic behavior collapses. 

To shed light on this behavior of the critical current as a function of the applied magnetic field, we perform TDGL simulations of both samples A and B. The results are presented in Fig.~\ref{fig:DevA} for sample A and in Supplemental Material for sample B~\cite{Supplemental}. They show discontinuous changes of the switching current of the nanobridge with each entry of vortices into the trap(s). The values of applied field at which vorticities of the boxes change, correspond to the experimental traces of Fig.~\ref{fig:ISW}, despite virtually no fitting of the parameters being used. When comparing the curves, one should bear in mind that the experimental data in Fig.~\ref{fig:ISW} presents the switching currents averaged over many testing pulses, with each pulse probing not necessarily the same vortex configuration, especially close to the transition field, when in the field-cooled sample various numbers of vortices can be trapped after application of the annealing pulse. On the other hand the theoretical calculation presented in Fig.~\ref{fig:DevA} is based on a single current-sweep for each value of the analyzed magnetic field. 

In the simulations, one sees that every new vortex entry in either box or the lead adjacent to the nanobridge changes the Cooper-pair density distribution at the bridge and thereby affects the switching current of the nanobridge. In Fig.~\ref{fig:DevA} one can distinguish separable branches in $I_{SW}(B_{\bot})$ dependence, each characterized by its own vortex number. The inspection of the order parameter (see insets in Fig.~\ref{fig:DevA}) reveals its enhancement on the edges of the boxes and in the nanobridge when a new vortex is added to the box, as arising from the suppression of the Meissner currents (cf. state with and without vortex in Video~\ref{fig:Video}). Consequently, it leads to the increase of the switching current at the transition regions. Once the vortices penetrate in the leads (in this case at fields beyond approximately equal to $27$~mT), the applied current may cause criticality in the leads before the criticality in the nanobridge is reached, leading to non-regular, but fully reproducible quasiexponential decay of $I_{SW}(B_\bot)$ dependence. This is clearly evidenced in both experimental and simulated data for both samples A and B.

Similar behavior is observed in samples C and D, albeit with decreased period of vortex entries and corresponding cusps in the $I_{SW}(B_\bot)$ characteristics, associated with larger vortex traps compared to samples A and B (the TDGL calculation for sample C is presented witching the Supplemental Material~\cite{Supplemental}). However, in the case of stripes connected by a nanobridge, where lateral trapping confinement is alleviated compared to the square boxes, the periodic behavior of $I_{SW}(B_{\bot})$ is lost - as shown for samples E and F in Fig.~\ref{fig:ISW}. We show the corresponding simulated data for sample E in Fig.~\ref{fig:DevE}, and note the very good agreement in both the slope of $I_{SW}(B_{\bot})$ and the penetration field of vortices in the leads (stripes). However, upon further increase of the magnetic field, one notices increasingly pronounced instability and the onset of criticality in leads in experimental data at significantly lower fields than in simulations. The width of the leads (approximately equal to $5\xi$) allows for the formation of just one vortex row \cite{Cordoba2019}, whose density is fast increasing with magnetic field, and whose stability is strongly affected by the quality of sample edges and the heating incurred during vortex entry and motion, which may explain our observations. For additional comparison, we show in the inset of Fig.~\ref{fig:DevE} the behavior of the critical current of the stripe alone, without the nanobridge (constriction), that once again validates the penetration field of vortices (the first cusp seen) and matches the tail characteristics of the simulated data for sample E at higher fields when the switching current of the nanostripe becomes lower than the switching current of the bridge.

\begin{video}[!tb]
\centering
\includegraphics[width=0.5\textwidth]{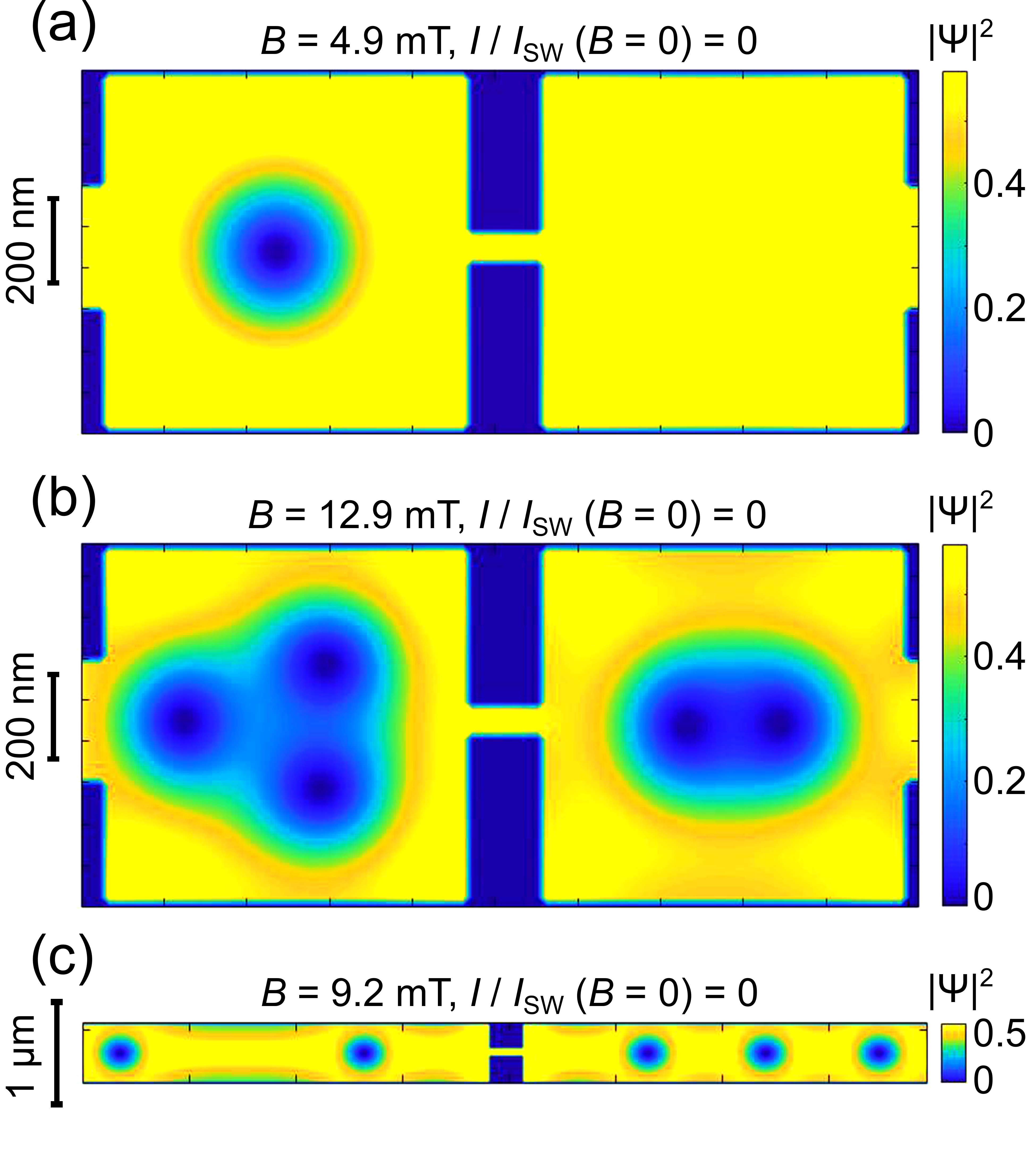}
\caption{\label{fig:Video}Visualization of the simulation results for (a) a single vortex and (b) one vortex out of five expelled from device A, and (c) a row of vortices expelled from device E at constant perpendicular magnetic fields and increasing applied currents.}
\end{video}

\subsection{High sensitivity at transitions between the vortex states}
At this stage, to emphasize both stability and sensitivity of the performed measurement, we detail the measured $I_{SW}(B_{\bot})$ trace in Fig.~\ref{fig:Zoom} for the sample with double trap and $W=860$~nm as a representative example (sample A, cf. Fig.~\ref{fig:ISW}; the analogous detailed traces for samples B, C and D are provided in Fig.~S5-S7 of the Supplemental Material~\cite{Supplemental}). One can see that at the transitions $I_{SW}$ is visibly enhanced (see e.g. the first and third vortex entries), but the increase is often preceded by a reduction of the switching current in a narrow range of applied magnetic field. We postulate that such behavior is a result of dissipation appearing in the trap when the testing pulse, when ramped up, expels the vortex before the switching current is reached. In such a case expelled vortex creates quasiparticles, whose presence decreases the switching current of the bridge. Such an expulsion is only possible for a narrow range of fields, since vortices are stronger bound in local energy minima further away from the transition fields - Bean-Livingston barrier grows with the field~\cite{Bean1964,Pekker2011,MorganWall2015}. Also, to observe the expulsion of a vortex from the box, the expulsion current must be lower than switching current of the bridge $I_{SW}$. This reasoning is in line with simulation data (Fig.~\ref{fig:DevA}): in general the entry of the additional vortex increases the value of switching current, but there are cases when a single vortex is pushed out from the box before the criticality in the bridge is reached [see Video~\ref{fig:Video}(a) or Video~\ref{fig:Video}(b)]. Vortex expulsion presumably creates more quasiparticles, reducing the value of the critical current in the bridge. Taking this interpretation, we expect that for the discussed sample, close to the transition from the Meissner state to the state with one vortex in each box, there is no vortex expulsion that would precede the switching process in the junction (transition no. 1 in Fig.~\ref{fig:Zoom}). Here, the switching current is enhanced due to the presence of vortices in the boxes. During transition no. 2, when one additional vortex is added to each box, there is small, but well-measurable suppression of the switching current. The trace is somewhat rounded which can be ascertained to statistical character of the measurement, which averages 1000 realizations for a given field: for some testing pulses vortex is expelled before criticality in the bridge is reached - such instances correspond to a suppressed value of the switching current, and for other testing pulses switching happens before vortex is expelled - such cases account for an enhanced value of the switching current. The picture is much clearer for transition no. 3, especially for positive fields, with a well-defined field value for the next vortex entry into one of the boxes, followed by the narrow region of magnetic fields where the switching current is enhanced due to the presence of this one additional vortex. The subsequently suppressed switching current corresponds to the magnetic field for which additional vortex enters into the second box. Unlike the earlier additional vortex, the second one can be expelled when ramping up the testing current, causing dissipation and reducing the critical current of the junction. However, at larger fields, its stability is enhanced and it can no longer be pushed away with the current lower than the switching current of the bridge. In that case, the presence of the second additional vortex increases the switching current even further, compared to the case with only one additional vortex. Similar analysis can be carried out for other transitions, with transition no. 5 looking especially appealing. Here, for negative magnetic fields, we see two overlapping clearly distinct ranges of magnetic fields, where the expulsion of one additional vortex is possible. The overlap corresponds to the simultaneous expulsion of the two additional vortices producing a pronounced dip in the switching current trace. The suppressed section of the trace to the right from the dip corresponds to the scenario with only one additional vortex in the box which can be expelled with the testing current. The suppressed part of the trace to the left from the dip corresponds to the scenario with two additional vortices in the boxes: one that can be expelled with the testing current and the second, which is now stable, although it has been mobile at lower magnetic field amplitudes. As intuitively expected, simultaneous expulsion of the two vortices should lead to stronger dissipation and larger reduction of the switching current, as compared to the case with only one vortex expelled. One should note that such effects are not considered in our simulations; our TDGL approach does not include thermal effects due to moving vortices and dissipation, so that the calculated switching current is only dependent on the vortex configuration at the switching threshold.

\section{Discussion}
For all samples with vortex traps we see a regular, ordered behavior of $I_{SW}$, while for those with  superconducting stripes we observe an intricate trace of $I_{SW}(B_{\bot})$ above vortex penetration field $B_0$. All measurements are fully reproducible [as shown in Fig.~\ref{fig:ISW}]. The detected behavior of vortex states as a function of trap size $W$ remains in agreement with the seminal measurements on aluminum superconducting discs~\cite{Geim1997}, as well as squares~\cite{Milosevic2009}, where similar vortex-related features in magnetization and transport are reported. Obviously, the number of possible vortex configurations in the boxes increases with the number of vortices in the trap. These configurations may have similar free energies as shown by the numerical calculations for the similar systems~\cite{Baelus2006,Moshchalkov2013}. We expect that some of the steps visible in our $I_{SW}(B_{\bot})$ characteristics may be attributed to the transitions between those states, as the switching current of the nanobridge for a particular vortex configuration will be sensitive to the proximity of a vortex to the bridge. 

Our reported sensitivity of the junction to the vicinal Meissner currents are in agreement with the findings of Ref.~\onlinecite{Timmermans2016}, where suppression of the zero bias conductance of the scanning tunneling tip is explained by a reduction of the supercurrents running at the sample edge when each next vortex enters the mesoscopic superconductor. Similar local restoration of the order parameter related to the adding of a new vortex was discussed in Ref.~\onlinecite{Kadowaki2004} and was detected as the increased voltage at a fixed current running through the normal-metal-insulator-superconductor (NIS) junction probing the edge of a mesoscopic disc. In all those cases vortex entry reduced the Meissner screening currents which in turn enhanced the superconducting gap at the geometrical borders. This directly affects the superconducting critical current of an attached nanobridge (as in our case) and changes the number of quasiparticles responsible for local charge transport (as in the aforementioned references \cite{Timmermans2016,Kadowaki2004}). 

In the structures studied in our experiments, the energetic minimum of the single vortex is most likely localized in the center of the traps. By nanoengineering the trap it is possible to pin the vortex closer to the bridge, thus enhancing its influence on the distribution of Meissner currents in the bridge. Such a pinning center can be favorably defined as a hole in the trap \cite{Bruyndoncx1999,Milosevic2007} or as a region of suppressed superconductivity, e.g. due to the nanoisland of copper or gold deposited on top of the trap~\cite{Timmermans2016}. We should probably emphasize that our samples exhibit some granulation (approximately $15-20$~nm in size, cf. Fig.~S1 within the Supplemental Material~\cite{Supplemental}) and some related pinning, which may affect the vortex configurations. Nevertheless the switching currents are to a large extent the same for positive and negative magnetic fields (cf. Fig.~\ref{fig:Zoom}). The small asymmetry of the results with respect to the field polarity may come from the unique shape and morphology of each vortex box and thus different configurations of the vortex-pinning ensemble. Consequently, vortices may occupy different positions when reversing the sign of the applied magnetic field or testing current.

Our device is compatible with current biases even at the level of a few hundreds of microamps i.e. in the range where expulsion of vortices should be possible, as suggested by experimentally measured suppression of the switching current (cf. Fig.~\ref{fig:Zoom} - transitions no. 3,4,5) and supported by simulation. It is a key feature distinguishing our arrangement from settings where tunneling Josephson junctions are used to probe the vortex matter in mesoscopic samples. As an illustration of such an effect, we show the time-dependent simulation results of the expulsion of a single vortex and one vortex out of five in device A [Supplemental Videos~\ref{fig:Video}(a) and ~\ref{fig:Video}(b), respectively], and the expulsion of a row of vortices from device E [Video~\ref{fig:Video}(c)] at a certain value of the external magnetic field and while ramping up the applied current. In the simplest scenario such an expulsion would allow switching currents of the nanobridge to be changed, offering a feature to build vortex-based logical devices\cite{Krasnov2015, Milosevic2009,Milosevic2010,melnikov_vinokur_2002}.

The presented study implicates that superconducting devices should be made with narrow wires to eliminate the existence of unwanted vortices that contribute to noise and decoherence of quantum circuits. The empirically found criterion for thin aluminum stripes offers $B_0=1.34\Phi_0/W^2$ as a threshold value for the first vortex entry.

\section{Conclusion}

We fabricate and thoroughly test several submicron superconducting devices in which the Dayem nanobridge as a sensing element is connected to aluminum nanosquare traps or stripes. We demonstrate that the pattern of the switching current of the nanobridge in applied magnetic field $I_{SW}(B_{\bot})$ is governed by configuration and dynamics of vortices in the adjacent traps. The Ginzburg-Landau simulations give insight into the behavior of the Meissner screening currents influenced by the arrangement of vortices and affecting the criticality of the nanobridge. Experimental data show additional intricate behavior, where dissipation associated with expulsion of a single vortex leads to pronounced suppression of the critical current of the nanobridge. Our device delivers a convenient \textit{in situ} probe for the presence and/or location of vortices in nanoengineered superconductors. Conversely, we show that transport properties of a Dayem bridge in magnetic field can be controlled by the size and vorticity of the adjacent islands. We anticipate that the presented approach is compatible with the manipulation of the vortex configuration with electric current and paves the way to demonstaration of a novel-concept logical devices.

\section*{Acknowledgments}
This work is supported by the Foundation for Polish Science project ``Stochastic thermometry with Josephson junction down to nanosecond resolution'' (First TEAM/2016-1/10), National Science Centre Poland project ``Thermodynamics of nanostructures at low temperatures'' (Sonata Bis-9, No. 2019/34/E/ST3/00432), the Research Foundation - Flanders (FWO), and the EU COST actions CA16218 NANOCOHYBRI and CA21144 SUPERQUMAP.

\end{document}